\documentclass[a4paper]{jpconf}
\usepackage{graphicx}
\usepackage{xspace}

\def\modell{{\sl poly-gonato} model\xspace}
\def\knee{{\sl knee}\xspace}
\def\knees{{\sl knees}\xspace}
\def\Xmax{$X_{max}$\xspace}
\def\gcm2{g/cm$^2$}
\def\eref#1{eq.~(\ref{#1})}
\def\fref#1{Fig.~\ref{#1}}
\def\sref#1{sect.~\ref{#1}}

\def\xx{\vspace*{-3mm}}

\begin{document}
\title{The end of the galactic cosmic-ray energy spectrum --- a
       phenomenological view%
       \footnote{Invited talk, presented at the "Workshop on Physics of the End
       of the Galactic Cosmic Ray Spectrum", Aspen, USA, April 25 - 29, 2005}}
\author{J\"org R. H\"orandel$^1$, N.N.Kalmykov$^2$, A.V.Timokhin$^3$}
\address{$^1$ Institute for Experimental Nuclear Physics, University of
         Karlsruhe, P.O. Box 3640, 76021 Karlsruhe, Germany}
\address{$^2$ Skobeltsyn Institute of Nuclear Physics, Lomonosov Moscow State
         University, Russia}
\address{$^3$ Faculty of Physics, Lomonosov Moscow State University, Russia}

\ead{hoerandel@ik.fzk.de}

\begin{abstract}
 Two structures in the all-particle energy spectrum of cosmic rays, the knee at
 4~PeV and the second knee around 400~PeV are proposed to be explained by a
 phenomenological model, the \modell,  connecting direct and indirect
 measurements.  Within this approach the knee is caused by a successive cut-off
 of the flux for individual elements starting with protons at 4.5~PeV. The
 second knee is interpreted as the end of the stable nuclei of the periodic
 table.  To check some key features of this model calculations of the cosmic
 ray energy spectrum and the propagation path length at energies from $10^{14}$
 to $10^{19}$~eV have been performed within the framework of a combined
 approach based on the diffusion model of cosmic rays and a direct simulation
 of charged-particle trajectories in the Galaxy. 
\end{abstract}

\section{Introduction}
The all-particle energy spectrum of cosmic rays follows power laws over many
orders of magnitude in energy. The flux decreases from values of several
1000~(m$^2$ sr s)$^{-1}$ at GeV energies to values below 0.01~(km$^2$ sr
a)$^{-1}$ at energies exceeding 100~EeV. At a closer look some structures are
seen in the spectrum.  In particular, the \knee around 4~PeV, the second \knee
at about 400~PeV, and the ankle above 4~EeV \cite{naganowatson,pg}.  In many
contemporary models for galactic cosmic rays the \knee is attributed to a
cut-off for the flux of light elements at energies of several PeV, while the
flux of heavier elements is supposed to continue to higher energies.  As
explanations the maximum energy attained during the acceleration process,
diffusive losses from the Galaxy, or interactions close to the source, during
propagation, as well as within the atmosphere are proposed \cite{origin}.

\begin{figure}[t]
  \includegraphics[width=0.49\textwidth]{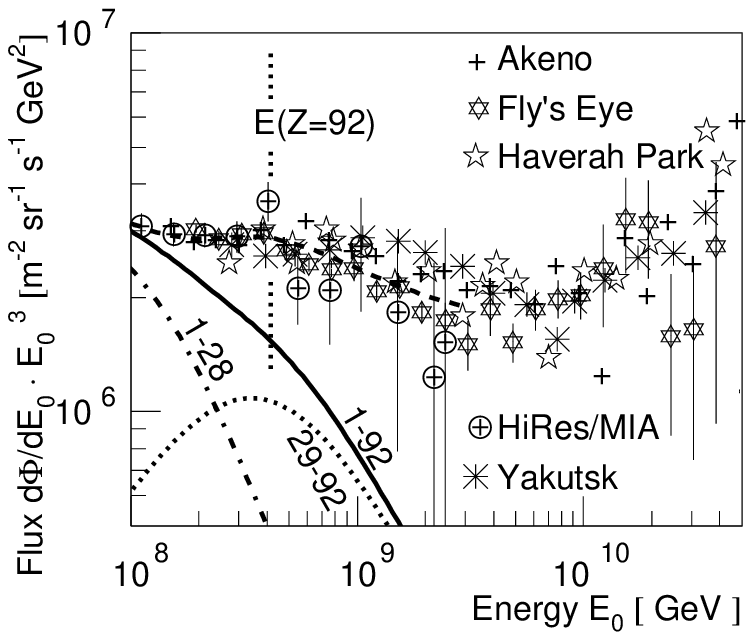}
  \includegraphics[width=0.49\textwidth]{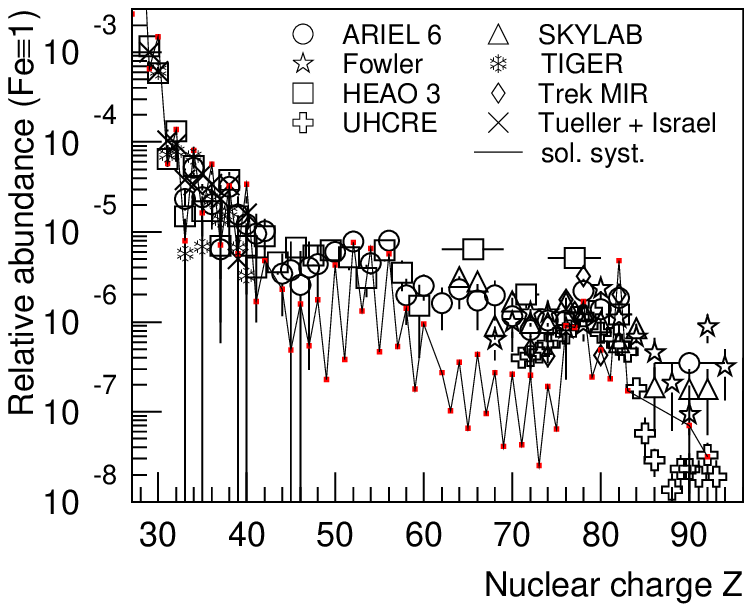}
  \xx
 \begin{minipage}[t]{18pc}
  \caption{\label{crflux} All-particle cosmic ray energy spectrum as
	   obtained by several air shower experiments in the region of the
	   second \knee, for references see \cite{pg}. The lines indicate
	   contributions for elemental groups with the given nuclear charge
	   range. The cut-off energy $92\cdot\hat{E}_p$ is indicated.}
 \end{minipage}\hspace{2pc}%
 \begin{minipage}[t]{18pc}
  \caption{\label{ariel} Relative abundance of cosmic-ray elements ($Z>28$)
	   normalized to Fe$\equiv 1$ from various experiments around 1~GeV/n.
	   For references see \cite{pg,tiger03}.  For comparison, abundances in
	   the solar system \cite{cameron} are presented as well, normalized to
	   Fe.}
 \end{minipage}
\end{figure}

The present work draws special attention on the second \knee at energies around
400~PeV. The all-particle energy spectrum as obtained by several experiments is
shown in \fref{crflux} in the energy region around the second knee.  A change
of the steepening of the average all-particle spectrum around 400~PeV is
visible.  In the following the possibility is explored, that this structure can
be interpreted as the end of the galactic cosmic-ray component.  It should be
pointed out that the position of the second \knee is found to be at about
$92\times\hat{E}_p=414$~PeV, where $\hat{E}_p=4.5$~PeV is the position of the
proton \knee \cite{pg}. In such a picture all stable elements known from the
periodic table of elements and found in cosmic rays are expected to contribute
to the galactic cosmic-ray component. A cut-off for the individual elemental
spectra for nuclei with charge number $Z$ at energies
$\hat{E}_Z=Z\cdot\hat{E}_p$ is anticipated.

The sketched picture is based on the \modell. Its principle ideas are outlined
in \sref{polygonato}. The model is a phenomenological approach connecting
direct and indirect measurements of cosmic rays in the energy range from 1~GeV
up to 1~EeV. The plausibility of some of its characteristic properties is
checked by a detailed treatment of the propagation of cosmic rays in the
Galaxy.  In \sref{difmod} the propagation of cosmic-ray particles in the
magnetic field of the Galaxy is investigated within a diffusion model.
Predictions are discussed for the energy spectrum, the path length in the
Galaxy, and the probability for a cosmic-ray particle to suffer nuclear
interaction during propagation. A summarizing discussion (\sref{discussion})
closes the article.

\section{The Poly-Gonato Model} \label{polygonato}
The basic idea of the \modell is to extrapolate the energy spectra as obtained
by direct measurements to higher energies and compare the sum of all elements
to the all-particle spectrum observed with air shower experiments \cite{pg}.
Within the model the measured energy spectra of individual elements are
parametrized assuming power laws and taking into account the solar modulation
at low energies. 

At energies in the GeV region the solar modulation of the energy spectra is
described using the parametrization
\begin{equation}\label{solmod}
 \frac{\textstyle d\Phi_Z}{\textstyle dE}(E)= 
     N \frac{\textstyle E(E+2 m_A)}{\textstyle E+M} \cdot
     \frac{\textstyle \left(E+M+780\cdot e^{-2.5\cdot10^4 E}\right)^{\gamma_Z}}
           {\textstyle E+M+2 m_A} ,
\end{equation}
adopted from \cite{bonino}.  $N$ is a normalization constant, $\gamma_Z$ the
spectral index of the anticipated power law at high energies (see below),
$E=E_0/A$ the energy per nucleon, $m_A$ the mass of the nucleus with mass
number $A$, and $M$ the solar modulation parameter.  A value $M=750$~MeV is
used for the parametrization.

\begin{figure}[t]
  \includegraphics[width=0.49\textwidth]{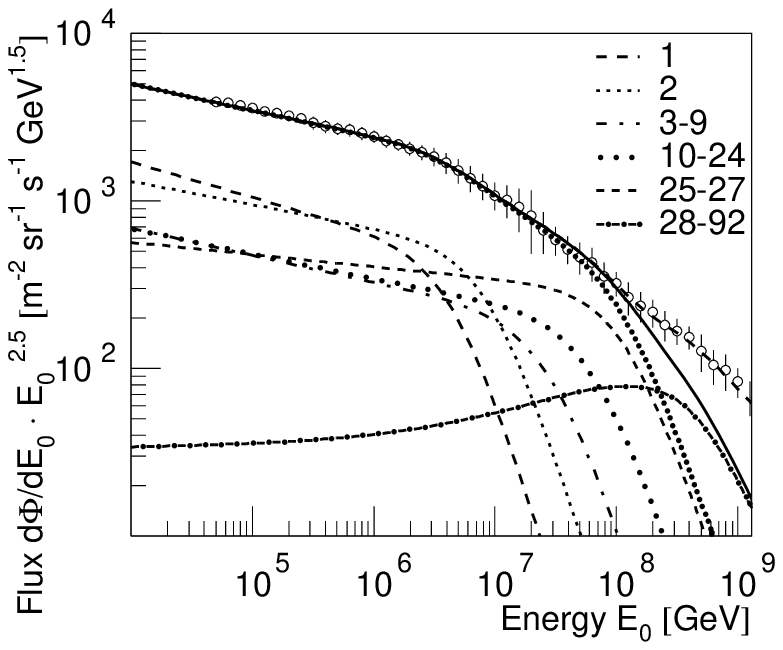}
  \includegraphics[width=0.49\textwidth]{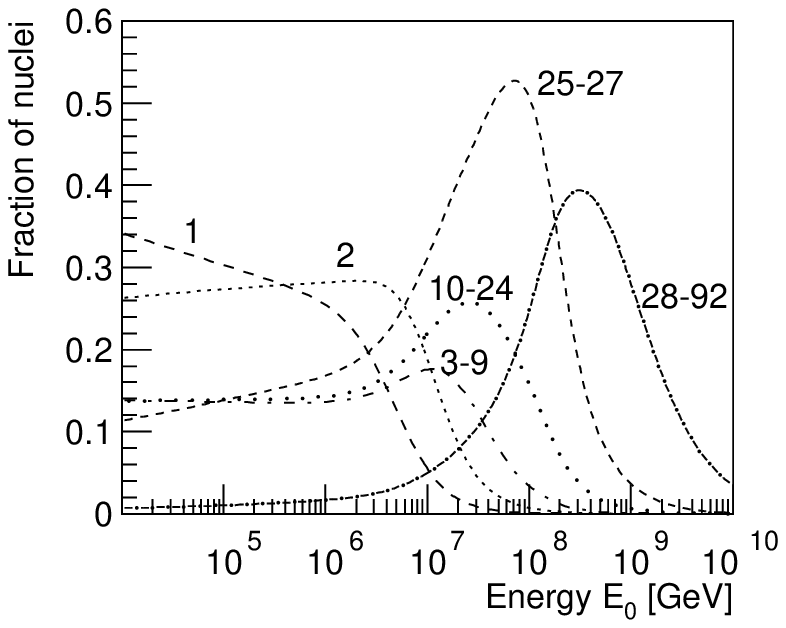}
  \xx
 \begin{minipage}[t]{18pc}
  \caption{\label{polyg}Cosmic ray energy spectra according to the \modell.
	   The data points represent the normalized all-particle flux obtained
	   by several experiments \cite{pg}.}
 \end{minipage}\hspace{2pc}%
 \begin{minipage}[t]{18pc}
  \caption{\label{uhfract}Relative fraction of elements with the indicated
           nuclear charge numbers according to the \modell.}
 \end{minipage}
\end{figure}

Above $Z\cdot 10$~GeV, the modulation due to the magnetic field of the
heliosphere is negligible and the energy spectra of cosmic--ray nuclei are
assumed to be described by power laws.  However, extrapolating the elemental
spectra with power laws, the all-particle spectrum obtained will overshoot the
measured all-particle spectrum above several PeV. Hence, a cut-off in the
spectra for the individual elements is introduced.  Such a cut-off is motivated
by various theories for the origin of the \knee \cite{origin}.  Inspired by such
theories the following ansatz is adopted to describe the energy dependence of
the flux for particles with charge $Z$ 
\begin{equation}\label{specfundg}
  \frac{d\Phi_Z}{dE_0}(E_0) = \Phi_Z^0 E_0^{\gamma_Z}
       \left[1+\left(\frac{E_0}{\hat{E}_Z}\right)^{\textstyle\epsilon_c}\right]
       ^\frac{\textstyle-\Delta\gamma}{\textstyle\epsilon_c}.
\end{equation}
The absolute flux $\Phi_Z^0$ and the spectral index $\gamma_Z$ quantify the
power law. The flux above the cut-off energy is modeled by a second and steeper
power law.  $\Delta\gamma$ and $\epsilon_c$ characterize the change in the
spectrum at the cut-off energy $\hat{E}_Z$.  Both parameters are assumed to be
identical for all elements, $\Delta\gamma$ being the difference in the spectral
indices below and above the respective \knees and $\epsilon_c$ describes the
smoothness of the transition from the first to the second power law. 
To study systematic effects also a common spectral index for all elements above
their respective \knee has been tried, see \cite{pg}.

Different relations for the cut-off energy $\hat{E}_Z$ have been checked.  The
investigations showed that a rigidity dependent cut-off with
$\hat{E}_Z=\hat{E}_p\cdot Z$, with the proton cut-off energy $\hat{E}_p$, fits
the data best \cite{pg}.  The all--particle spectrum is obtained by summation
of the flux $d\Phi_Z/dE_0(E_0)$ for all cosmic--ray elements.

The absolute flux and the spectral indices for the elements from protons to
nickel are derived from direct measurements above the atmosphere.  For the
heavier elements only abundances are known at energies around 1~GeV/n. In
\fref{ariel} the relative abundance normalized to Fe $\equiv1$ is presented as
function of the nuclear charge number. One recognizes that all elements up to
the end of the periodic table are present.  At energies of 1~GeV/n the measured
abundances are heavily suppressed by the heliospheric modulation of cosmic rays
--- this is taken into account in the \modell using \eref{solmod}.  In
addition, the measured abundances are extrapolated to high energies using an
empirical relation for the spectral indices $-\gamma_Z= A + B\cdot Z^C$ with
the values $A=2.70\pm0.19$, $B=(-8.34\pm4.67)\cdot10^{-4}$, and
$C=1.51\pm0.13$.

The average all-particle flux obtained by several air shower experiments is
given in \fref{polyg} by the data points. A fit to these points yields the
parameters for the \modell $\hat{E}_p=4.49\pm0.51$~PeV,
$\Delta\gamma=2.10\pm0.24$, and $\epsilon_c=1.90\pm0.19$.  The expected spectra
for elemental groups corresponding to these parameters are shown in
\fref{polyg}.  The all-particle spectrum obtained by summation of all elements
fits the experimental values well. It can be recognized that within this
approach the \knee is a consequence of the subsequential cut-offs for the
individual elemental spectra, starting with protons at 4.5~PeV.  The shape of
the all-particle spectrum above this energy is determined by the overlay of the
individual cut-offs for all elements. The second knee is related to the
cut-offs for the heaviest elements at the end of the periodic table or the end
of the galactic component.

The relative contributions of elemental groups to the all-particle spectrum are
presented in \fref{uhfract}.  The most dominant group is the iron group
($Z=25-27$), at energies around 70~PeV more than 50\% of the all-particle flux
consists of these elements.  Ultra heavy elements ($Z\ge28$) are expected to
contribute at 300~PeV with slightly less than 40\% to the all-particle flux.

\begin{figure}[t]
 \includegraphics[width=18pc]{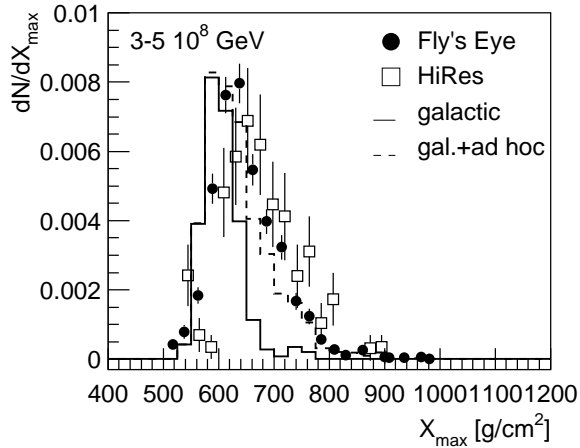}\hspace{2pc}%
 \begin{minipage}[b]{18pc}
  \caption{\label{xmax}Distribution of the depth of the shower maximum \Xmax
	   measured by the Fly's Eye \cite{flyseyexmax} and HiRes
	   \cite{hiresxmax} experiments.  The measured values are compared with
	   simulated results using QGSJET with lower cross sections \cite{wq}.
	   Galactic component according to the \modell (solid line) and
	   galactic plus {\sl ad-hoc} component (dashed line), see also \cite{wq}.}
 \end{minipage}
\end{figure}

Around 300~PeV measurements of the average depth of the shower maximum (\Xmax)
are available from the Fly's Eye and HiRes experiments, see \fref{xmax}. These
data are used to check the predictions of the \modell.  With the air shower
simulation program CORSIKA \cite{corsika} and a modified version of the
hadronic interaction model QGSJET with lower cross sections \cite{wq} the
expected \Xmax distribution has been calculated.  The composition of galactic
cosmic rays has been assumed according to the \modell.  The resulting
distribution is given in \fref{xmax} as solid histogram.  It should be
emphasized that this is not a fit to the measured \Xmax values, instead, the
fluxes are taken as {\sl predicted} by the model.  As can be inferred from the
figure, the left-hand side of the \Xmax distribution is represented quite
reasonably.  An {\sl ad-hoc} component has been introduced to describe also the
right-hand part of the distribution, for details see \cite{wq}.

It should be pointed out that the three parameters of the model, namely
$\hat{E}_p$, $\epsilon_c$, and $\Delta\gamma$ in \eref{specfundg} have been
determined by a fit to the {\sl all particle} spectrum derived from indirect
measurements.  That means the flux values for {\sl individual elements} in the
region of their cut-off are real predictions of the model. A comparison of the
predicted fluxes with the energy spectra for groups of elements as recently
derived from air shower measurements shows a quite satisfactory compatibility
between the \modell and the measurements \cite{origin}.

Two questions may rise when inspecting \fref{polyg}: Why are the spectra of
heavy elements flatter than the spectra for light elements? And: can there be a
sufficient contribution of ultra heavy elements at energies around $10^8$~GeV?
To obtain a quantitative estimate on this items in the following section the
propagation of cosmic rays in the Galaxy is discussed.

\section{Propagation of Cosmic Rays in the Galaxy} \label{difmod}

To study cosmic-ray propagation in the Galaxy, a detailed knowledge of the
structure of the magnetic fields is essential. Unfortunately, the question
about the configuration of the galactic magnetic field remains open ---
different models exist based on experimental data
\cite{berezinsky,ruzmaikin,ptuskin,gorchakov}.  How cosmic rays are accelerated
to extremely high energies is another unanswered question. Although the popular
model of cosmic-ray acceleration by shock waves in the expanding shells of
supernovae (see e.g.  \cite{ellison,berezhko,sveshnikova}) is almost recognized
as "standard theory," there are still a number of unresolved problems.
Furthermore, the question about other acceleration mechanisms is not quite
clear, and could lead to different cosmic-ray energy spectra at the sources
\cite{berezinsky}.  

Different concepts are verified by the calculation of the primary cosmic-ray
energy spectrum, making assumptions on the density of cosmic-ray sources, the
energy spectrum at the sources, and the configuration of the galactic magnetic
fields. The diffusion model can be used in the energy range $E<10^{17}$~eV,
where the energy spectrum is calculated using the diffusion equation for the
density of cosmic rays in the Galaxy. At higher energies this model ceases to
be valid, and it becomes necessary to carry out numerical calculations of
particle trajectories for the propagation in the magnetic fields of the Galaxy.
This method works best for the highest energy particles, since the time for the
calculations required is inversely proportional to the particle energy.
Therefore, the calculation of the cosmic-ray spectrum in the energy range
$10^{14}-10^{19}$~eV has been performed within the framework of a combined
approach, the use of a diffusion model and the numerical integration of
particle trajectories.

\subsection{Assumptions} 
High isotropy and a comparatively long retention of cosmic rays in the Galaxy
($\sim10^7$ years for the disk model) reveal the diffusion nature of particle
motion in the interstellar magnetic fields. This process is described by
a corresponding diffusion tensor \cite{berezinsky,ptuskin,kalmykov}.  The
steady-state diffusion equation for the cosmic-ray density $N(r)$ is
(neglecting nuclear interactions and energy losses)
\begin{equation} \label{diffeq}
 - \nabla_iD_{ij}(r)\nabla_jN(r)=Q(r) .
\end{equation}
$Q(r)$ is the cosmic-ray source term, $D_{ij}(r)$ is the diffusion tensor.

Under the assumption of azimuthal symmetry and taking into account the
predominance of the toroidal component of the magnetic field, \eref{diffeq} is
presented in cylindrical coordinates as 
\begin{equation} \label{zylfun}
\left[ 
  -\frac{1}{r}\frac{\partial}{\partial r} r D_\perp \frac{\partial}{\partial r} 
  -           \frac{\partial}{\partial z}   D_\perp \frac{\partial}{\partial z} 
  -           \frac{\partial}{\partial z}   D_A     \frac{\partial}{\partial r} 
  +\frac{1}{r}\frac{\partial}{\partial r} r D_A     \frac{\partial}{\partial z} 
\right] N(r,z)=Q(r,z) ,
\end{equation}
where $N(r,z)$ is the cosmic-ray density averaged over the large-scale
fluctuations with a characteristic scale $L\sim100$~pc \cite{ptuskin}.
$D_\perp\propto E^m$ is the diffusion coefficient, where $m$ is much less than
one ($m\approx0.2$), and $D_A\propto E$ the Hall diffusion coefficient.  The
influence of Hall diffusion becomes predominant at sufficiently high energies
($>10^{15}$~eV). The sharp enhancement of the diffusion coefficient leads to
the excessive cosmic-ray leakage from the Galaxy at energies $E>10^{17}$~eV.
For investigating the cosmic-ray propagation at such energies it is necessary
to carry out calculations of the trajectories for individual particles.  

The numerical calculation of trajectories is based on the solution of the
equation of motion for a charged particle in the magnetic field of the Galaxy.
The calculation was carried out using a fourth order Runge-Kutta method.
Trajectories of cosmic rays were calculated until they left the Galaxy.
Testing the differential scheme used, it was found that the accuracy of the
obtained trajectories for protons with an energy $E=10^{15}$~eV after passing a
distance of 1~pc amounts to $5\cdot10^{-8}$~pc.  The retention time of a proton
with such an energy averages to about 10 million years, hence, a total error
for the trajectory approximation by the differential scheme is about 0.5~pc.

The magnetic field of the Galaxy consists of a large-scale regular and a
chaotic component $\vec{B}=\vec{B}_{reg}+\vec{B}_{irr}$.  A purely azimuthal
magnetic field was assumed for the regular field 
\begin{equation}
 B_z=0, \quad B_r=0, \quad B_\phi=1~\mu\mbox{G} \cdot 
       \exp\left(-\frac{z^2}{z_0^2}-\frac{r^2}{r_0^2}\right) ,
\end{equation}
where $z_0=5$~kpc and $r_0=10$~kpc are constants \cite{ptuskin}.  The irregular
field was constructed according to an algorithm used in \cite{zirakashvili},
that takes into account the correlation of the magnetic field intensities in
adjacent cells.  The radius of the Galaxy is assumed to be 15~kpc and the
galactic disk has a half-thickness of 200~pc. The position of the Solar system
was defined at $r=8.5$~kpc, $\phi=0^\circ$, and $z=0$~kpc. A radial
distribution of supernovae remnants along the galactic disk was considered as
sources \cite{kodaira}.

\subsection{Results}

\begin{figure}[t]
  \includegraphics[width=0.49\textwidth]{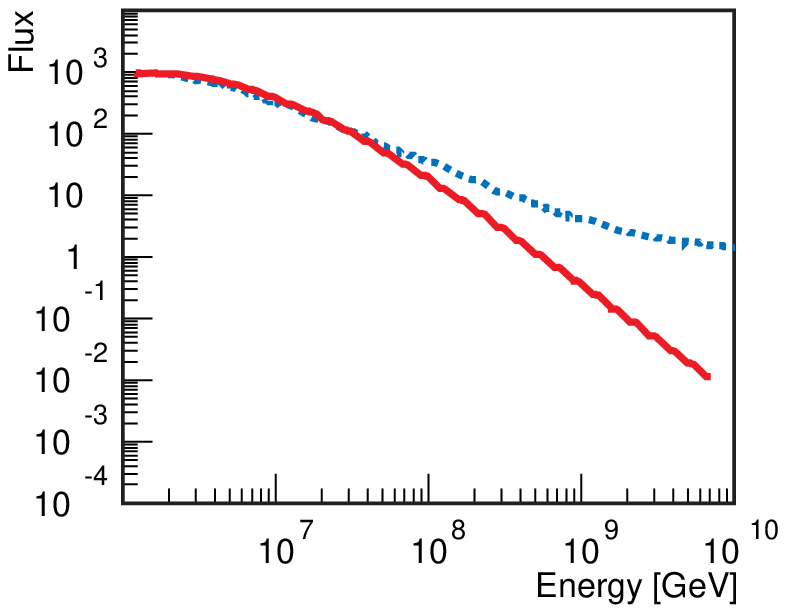}
  \includegraphics[width=0.49\textwidth]{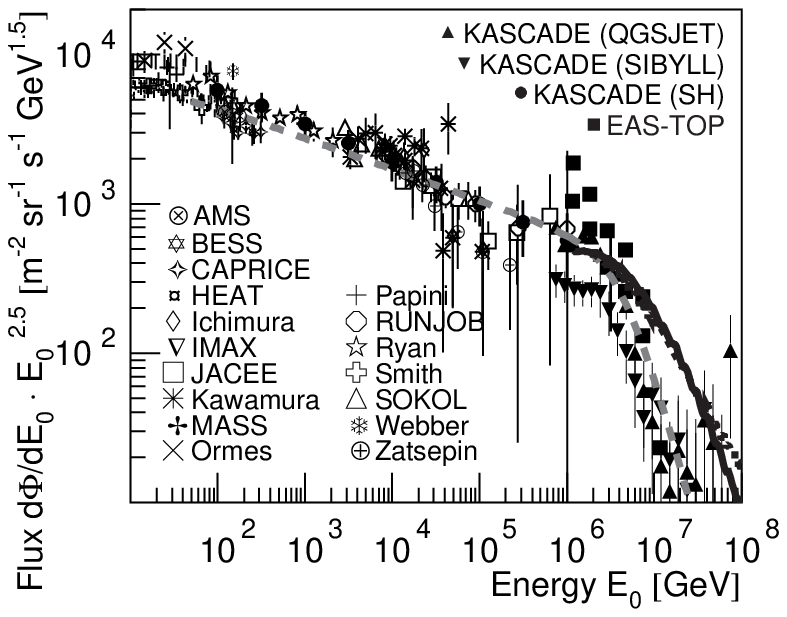}
  \xx
 \begin{minipage}[t]{18pc}
  \caption{\label{pspec}Calculated spectra for protons for the diffusion model
	   (solid line) and the numerical trajectory calculations (dotted
	   line).}
 \end{minipage}\hspace{2pc}%
 \begin{minipage}[t]{18pc}
  \caption{\label{pflux} Proton flux as obtained from various measurements, for
	   references see \cite{vulcano}, compared to the spectra shown
	   in \fref{pspec} and the \modell (dashed line).}
 \end{minipage} 
\end{figure}

The results for the calculations of the cosmic-ray proton spectrum are
presented in \fref{pspec}. These results were obtained using the diffusion
model and numerical calculations of trajectories. It is evident from the graph
that both methods give identical results up to about $3\cdot10^7$~GeV. At
higher energies there is a continuous decrease of the intensity in the
diffusion spectrum, which corresponds to the excessive increase in the
diffusion coefficient that leads to a large leakage of particles from the
Galaxy.  

An energy of $10^{17}$~eV can be defined as the conventional boundary to apply
the diffusion model. At this energy the results obtained with the two methods
differ by a factor of two and for higher energies the diffusion approximation
becomes invalid.

The predicted spectra are compared to direct and indirect measurements of the
primary proton flux in \fref{pflux}. In the depicted range there is almost no
difference between the two approaches. The relatively steep decrease of the
flux at energies exceeding 4~PeV is not reflected. On the other hand, the data
are described reasonably well by the \modell, as also shown in the figure.  The
observed change in the spectral index $\Delta\gamma\approx2.1$ according to the
\modell has to be compared to the value predicted by the diffusion model. In
the latter the change should be $1-m\approx0.8$ \cite{ptuskin}. The observed
value is obviously larger, which implies that the remaining change of the
spectral shape should be caused by a change of the spectrum at the source, e.g.
due to the maximum energy attained in the acceleration process.

The maximum energy and, therefore, the energy at which the spectrum steepens
depends on the intensity of the magnetic fields in the acceleration zone and on
a number of assumptions for the feedback of cosmic rays to the shock front.
The uncertainty of the parameters yields variations in the maximum energy
predicted by different models up to a factor of 100 \cite{origin,berezhko}.
Thus, there is no consensus about what the "standard model" is considered to
predict. For the time being, it is difficult to make definite conclusions from
the comparison between the experimental spectra for different elemental groups
and the "standard model" of cosmic ray acceleration at ultra high energies.  

\begin{figure}[t]
  \includegraphics[width=0.49\textwidth]{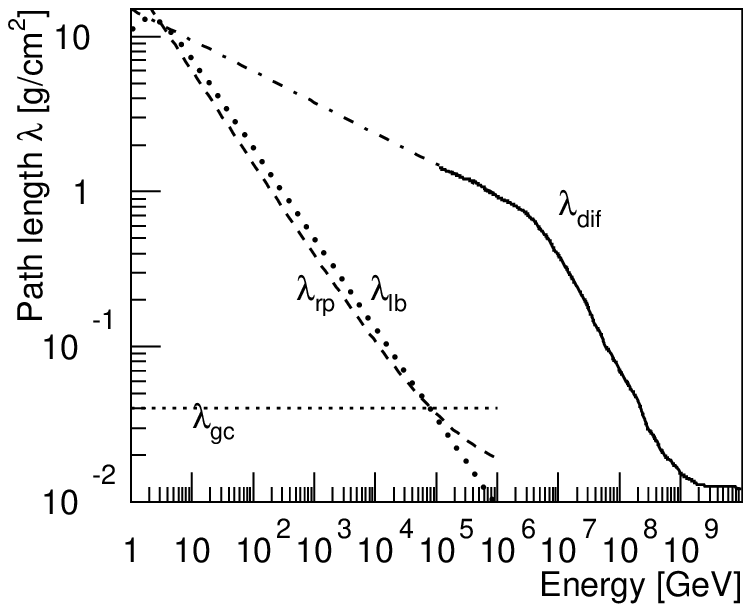}
  \includegraphics[width=0.49\textwidth]{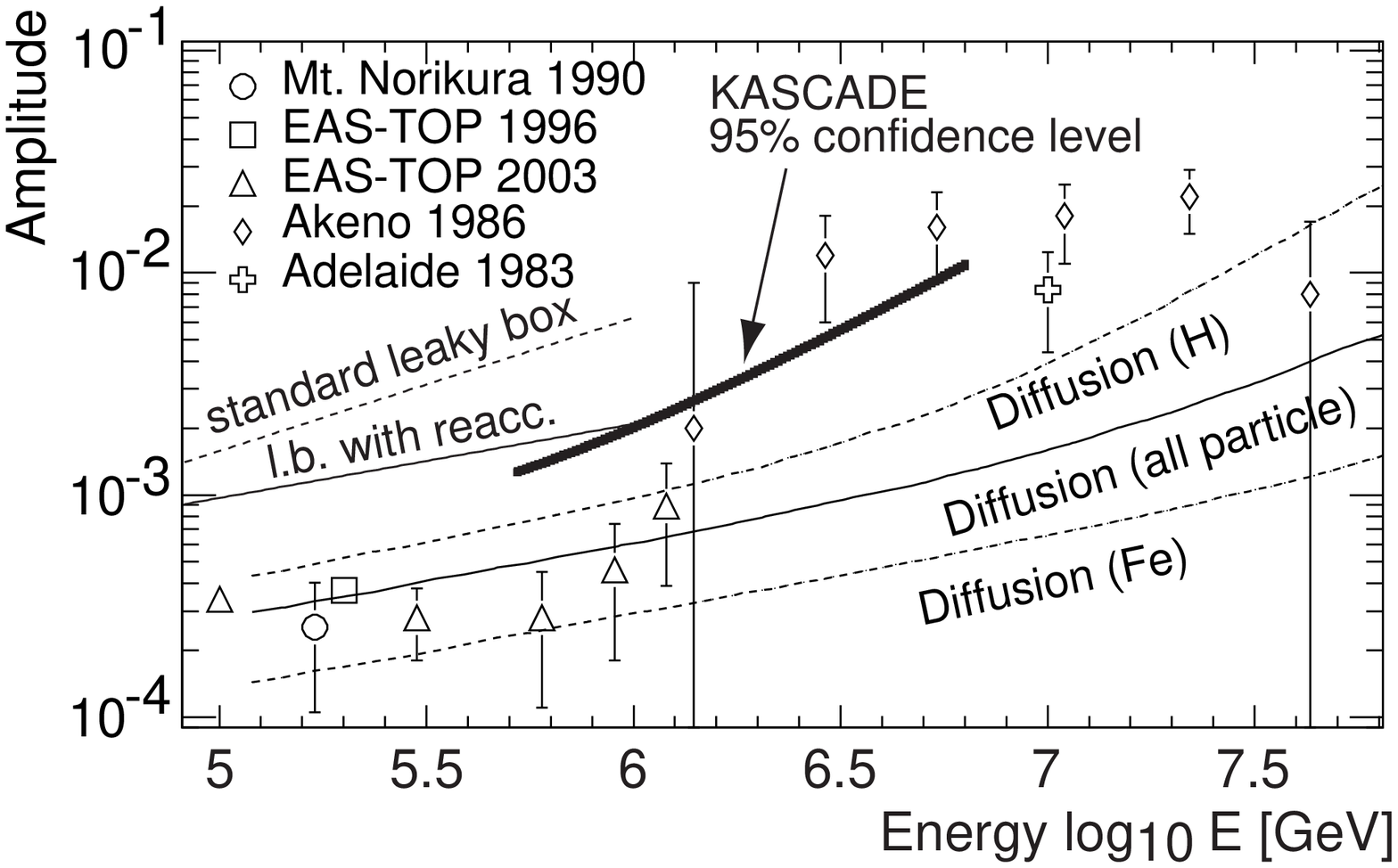}
  \xx
 \begin{minipage}[t]{18pc}
  \caption{\label{path length} Path length in the Galaxy for protons.
	   The values for the diffusion model ($\lambda_{dif}$) are indicated
	   by the solid line. They are extrapolated to lower energies by the
	   dashed dotted line.  The dashed and dotted lines indicate a
	   leaky-box model ($\lambda_{lb}$, \eref{leakyboxfun}) and a residual
	   path length model ($\lambda_{rp}$, \eref{swordyfun}). The dotted
	   line indicates the matter accumulated along a straight line from the
	   galactic center to the solar system ($\lambda_{gc}$).}
 \end{minipage}\hspace{2pc}%
 \begin{minipage}[t]{18pc}
  \caption{\label{aniso}Rayleigh amplitudes as function of energy for various
	   experiments, for references see \cite{kascade-aniso}. Additionally,
	   model predictions for leaky-box models \cite{ptuskinaniso} and a
	   diffusion model \cite{candiaaniso} are shown. The lines indicate the
	   expected anisotropy for primary protons, iron nuclei, and all
	   particles.}
 \end{minipage} 
\end{figure}

The obtained path length ($\lambda_{dif}$) in the Galaxy for protons as
function of energy is presented in \fref{path length}. The interstellar matter
density was taken as $n_d=1$~cm$^{-3}$ for the galactic disk and
$n_h=0.01$~cm$^{-3}$ for the halo.  For heavier nuclei with charge $Z$ the path
length scales with the rigidity, i.e.  is related to the values for protons
$\lambda(E)$ as $\lambda(E,Z)=\lambda(E/Z)$.  At the corresponding knees, the
amount of traversed material is less than 1~g/cm$^2$.  The dashed dotted line
indicates a trend at lower energies according to $\lambda\propto E^{-\delta}$.
To reach values around 10~g/cm$^2$ as obtained around 1~GeV, see below, one
needs a relatively small slope $\delta\approx0.2$ --- much lower than the value
usually assumed $\delta\approx0.6$.

Measurements of the ratio of secondary to primary cosmic-ray nuclei at GeV
energies are successfully described using a leaky-box model. For example,
assuming the escape path length as
\begin{equation} \label{leakyboxfun}
 \lambda_{lb}=\frac{26.7\beta ~\mbox{g/cm}^2}
              {(\beta R/1.0~\mbox{GV})^{0.58} + (\beta R/1.4~\mbox{GV})^{-1.4}}
\end{equation}
various secondary to primary ratios obtained by the ACE/CRIS and HEAO-3
experiments can be described consistently in the energy range from $\sim70$~MeV
to $\sim30$~GeV \cite{cris-time}.
A similar approach is the residual path length model \cite{swordy}, assuming the
relation
\begin{equation} \label{swordyfun}
 \lambda_{rp} = \left[ 6.0 \cdot 
     \left(\frac{R}{10~\mbox{GV}}\right)^{-0.6} + 0.013 \right] ~\mbox{g/cm}^2
\end{equation}
for the escape path length. Both examples are compared to the predictions of
the diffusion model in \fref{path length}.

\begin{figure}[t]
  \includegraphics[width=0.49\textwidth]{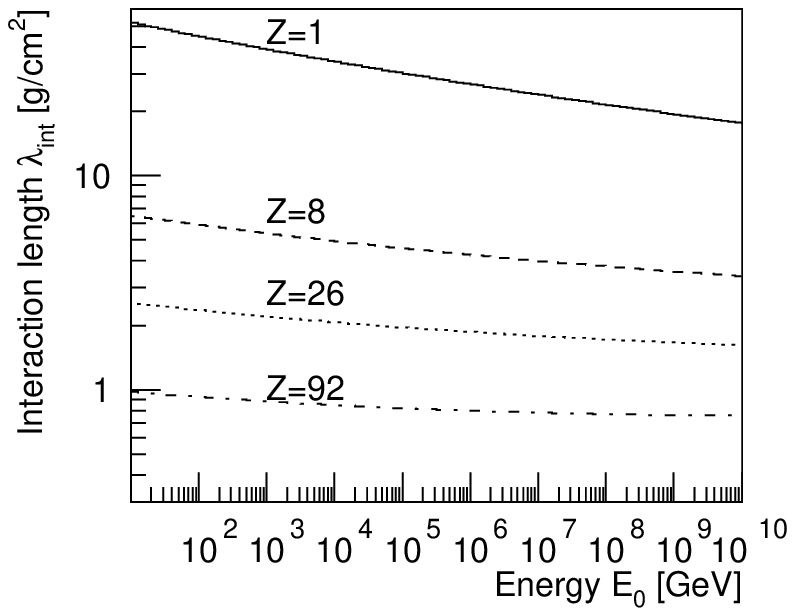}
  \includegraphics[width=0.49\textwidth]{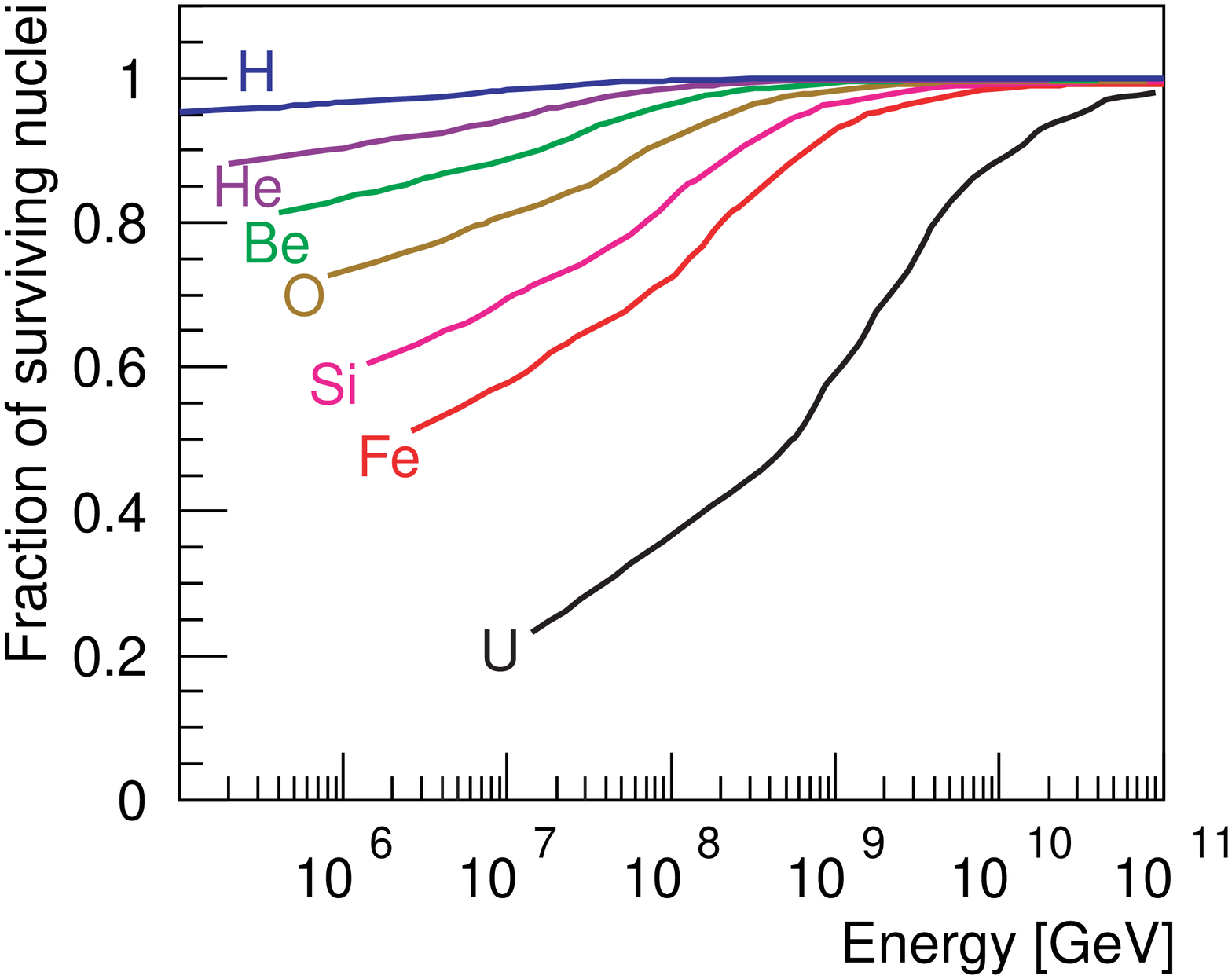}
  \xx
 \begin{minipage}[t]{18pc}
  \caption{\label{lambdaint}Interaction length for different elements based on
	   cross-sections according to the hadronic interaction model QGSJET,
	   the respective nuclear charge number is indicated.}
 \end{minipage}\hspace{2pc}%
 \begin{minipage}[t]{18pc}
   \caption{\label{fragment}Fraction of nuclei surviving without interaction in
   the Galaxy for different elements.}
 \end{minipage} 
\end{figure}

Extrapolating these relations to higher energies, the strong dependence of the
path length on energy ($\propto E^{-0.6}$) leads to extremely small values at PeV
energies. Above $10^5$~GeV the traversed matter would be less than the matter
accumulated along a straight line from the galactic center to the solar system
$\lambda_{gc}= 8~\mbox{kpc}\cdot 1~\mbox{proton/cm}^3 \approx
0.04~\mbox{g/cm}^2$. This value is indicated in \fref{path length} as dotted
line.

A similar conclusion can be derived from anisotropy measurements.  Rayleigh
amplitudes observed by different experiments are compiled in \fref{aniso}
\cite{kascade-aniso}.  Leaky-box models, with their extremely steep decrease of
the path length $\lambda\propto E^{-0.6}$, yield relative large anisotropies
even at modest energies, which seem to be ruled out by the measurements.  Two
versions of a leaky-box model \cite{ptuskinaniso}, with and without
reacceleration are represented in the figure. On the other hand, a diffusion
model \cite{candiaaniso}, which is based on the same basic idea \cite{ptuskin}
as the present work, seems to be compatible with the measured data.  For this
model the expected Rayleigh amplitudes are given for primary protons and iron
nuclei, as well as for a mixture of all elements.

Using nuclear cross sections according to the hadronic interaction model QGSJET
\cite{qgsjet} and assuming a number density $n_p=1$~proton/cm$^3$ in the
galactic disk, the interaction length of nuclei has been calculated.  The
results for four different elements are presented in \fref{lambdaint}.  The
values decrease slightly as function of energy.  Values for protons are in the
range $20-55$~\gcm2, the values decrease as function of nuclear charge and reach
values $<1$~\gcm2 for the heaviest elements.

The interaction probability for different nuclei has been calculated based on
the obtained path length and interaction parameters according to the QGSJET
model. Nuclear fragmentation is taken into account in an
approximate approach \cite{fragm}. It should be pointed out that a nuclear
fragment  conserves the trajectory direction of its parent if $Z/A$ 
in question is the same as for the primary nucleus and for most stable
nuclei the ratio $Z/A$ is close to $1/2$. The resulting fraction of nuclei
which survive without an interaction is presented in \fref{fragment} for
selected elements. 
It turns out that at the respective knees ($Z \cdot 4.5$~PeV) more than about
50\% of the nuclei survive without interactions, even for the heaviest
elements. This is an important result, since the \modell relates the
contribution of ultra-heavy cosmic rays to the second knee in the all-particle
spectrum around 400 PeV.  It should be noted that the fraction of surviving
nuclei would be even larger for a leaky-box model, with its low path length at
such energies.

\section{Discussion} \label{discussion}
The energy spectrum of cosmic rays at their source $Q(E)$ is related
to the observed values at Earth $N(E)$ as
\begin{equation}
N(E) = Q(E) 
  \left( \frac{1}{\lambda_{esc}(E)} + \frac{1}{\lambda_{int}(E)}\right)^{-1}
\end{equation}
with the propagation path length $\lambda_{esc}$ and the interaction length
$\lambda_{int}$. The corresponding values are presented in \fref{path length}
and \fref{lambdaint}, respectively.  It is frequently assumed that the
propagation path length decreases as function of energy $\lambda_{lb}\propto
E^{-0.6}$, as discussed above.  Since the interaction length is almost
independent of the primary energy this necessitates a spectrum at the sources
$Q(E)\propto E^{-2.1}$ to explain the observed spectrum at Earth $N(E)\propto
E^{-2.7}$.  However, the model by Berezhko \etal \cite{berezhko} predicts even
flatter spectra at the sources before the \knee and an even stronger dependence
of $\lambda_{esc}$ on energy is needed.
Recent measurements of the TeV $\gamma$ ray flux from a shell type supernova
remnant yield a spectral index $\gamma=-2.19\pm0.09\pm0.15$ \cite{hesssnr} in
agreement with the "standard model". However, e.g. for the Crab Nebula a
steeper spectrum with $\gamma=-2.57\pm0.05$ has been obtained \cite{hesscrab},
indicating that not all sources exhibit the same behaviour.

As has been discussed above, the dependence of the propagation path length
$\lambda_{esc}\propto E^{-0.6}$ can not be extrapolated to \knee energies.
Taking a value $\lambda_{esc}\propto E^{-0.2}$ as discussed in relation with
\fref{path length} necessitates additional assumptions on the spectral shape
$Q(E)$ at the source in order to explain the observed spectra with spectral
indices in the range $-\gamma \approx 2.55 - 2.75$ \cite{pg}.

Direct measurements seem to indicate that the spectra of light elements are
flatter as compared to heavy elements \cite{pg}.  The values of $\lambda_{int}$
for protons are at all energies larger than the escape path length
$\lambda_{dif}$. Hence, for protons the escape from the Galaxy is the dominant
process influencing the shape of the observed energy spectrum.  For iron nuclei
at low energies hadronic interactions are dominating
($\lambda_{int}\approx2~\mbox{g/cm}^2 < \lambda_{dif}$) and leakage from the
Galaxy becomes important at energies approaching the iron knee ($26\times
\hat{E}_p$).  However, for elements heavier than iron the interaction path
length is smaller than the escape path length for all energies, except above
the respective \knees.  At low energies the propagation path length exceeds
the interaction path length by about an order of magnitude.  For these elements
nuclear interaction processes are dominant for the shape of the observed
spectrum.  This may explain why energy spectra for heavy elements should be
flatter as compared to light nuclei. In particular, the ultra-heavy elements
suffer significantly from interactions at low energies.

In {\bf summary},
the results obtained show the effectiveness of the combined method to calculate
the cosmic-ray spectrum using a numerical calculation of trajectories and a
diffusion approximation. 
The calculated dependence of the propagation path length on energy suggests that
the difference between the predicted spectral index at the source
($\gamma\approx-2.1$) in the "standard model" and the experimental value
($\gamma\approx-2.7$) can not be explained by the energy dependence of the
escape path length solely.
The compatibility of the observed cosmic-ray energy spectrum with the "standard
model" requires additional assumptions.  
Most likely, it can be concluded that the \knee in the energy spectrum of cosmic
rays has its origin in both, acceleration and propagation processes.

In the \modell the \knee in the energy spectrum at 4.5~PeV is caused by a
cut-off of the light elements and the spectrum above the \knee is determined by
the subsequent cut-offs of all heavier elements at energies proportional to
their nuclear charge number. The second \knee around $400~\mbox{PeV} \approx
92\cdot\hat{E}_p$ is due to the cut-off of the heaviest elements in galactic
cosmic rays. Considering the calculated escape path length and nuclear
interaction length within the diffusion model, it seems to be reasonable that
the spectra for heavy elements are flatter as compared to light elements. The
calculations show also that even for the heaviest elements at the respective
\knee energies more than 50\% of the nuclei survive the propagation process
without interactions. This may explain why ultra-heavy elements are expected
to contribute significantly ($\sim 40\%$) to the all-particle flux at energies
around 400~PeV.

\section*{Acknowledgements}
The authors are grateful to J.~Engler, A.I.~Pavlov, and V.N.~Zirakashvili for
useful discussions.
N.N.K. and A.V.T. acknowledge the support of the RFBR (grant 05-02-16401).

\section*{References}

\end{document}